\documentstyle[12pt,aasms4]{article}
\def\alph{$\alpha$}
\def\be{\begin{equation}}
\def\bea{\begin{eqnarray}}
\def\ee{\end{equation}}
\def\eea{\end{eqnarray}}
\def\lam{$\lambda$}
\def\mic{$\mu$m}
\def\noi{\noindent}
\def\nsh{$n$(shell)}
\def\qa{$Q_{\rm ext}/x^{0.5}$}
\def\qex{$Q_{\rm ext}$}

\begin{document}

\title{Calculating Cross Sections of Composite Interstellar Grains}

\author{Nikolai V. Voshchinnikov\footnote{
Sobolev Astronomical Institute, Univ. of St. Petersburg;
Bibliotechnaya Pl. 2, Stary Peterhof;
St. Petersburg, Russia 198904; nvv@aispbu.spb.su}
and John S. Mathis\footnote{Dept. of Astronomy, Univ. of Wisconsin;
475 N. Charter St., Madison WI 53706, USA;
mathis@uwast.astro.wisc.edu}}

\begin{abstract}
Interstellar grains may be composite collections of particles of
distinct materials, including voids, agglomerated together. We
determine the various optical cross sections of such composite grains,
given the optical properties of each constituent, using an approximate
model of the composite grain. We assume it consists of many concentric
spherical layers of the various materials, each with a specified
volume fraction. In such a case the usual Mie theory can be generalized
and the extinction, scattering, and other cross sections determined
exactly.

We find that the ordering of the materials in the layering makes some
difference to the derived cross sections, but averaging over the
various permutations of the order of the materials provides rapid
convergence as the number of shells (each of which is filled by all of
the materials proportionately to their volume fractions) is
increased. Three shells, each with one layer of a particular
constituent material, give a very satisfactory estimate of the average
cross section produced by larger numbers of shells.

We give the formulae for the Rayleigh limit (small size parameter) for
multi-layered spheres and use it to propose an ``Effective Medium
Theory'' (EMT), in which an average optical constant is
taken to represent the ensemble of materials.

Multi-layered models are used to compare the accuracies of several
EMTs already in the literature. EMTs are worse for predicting
scattering cross sections than extinction, and considerably worse for
predicting $g$, the mean cosine of the angle of scattering. However,
the angular distribution of the scattered radiation depends
sensitively on the assumed grain geometry and should be taken with
caution for any grain theory. Our computation is vastly simpler than
discrete multipole calculations and may be easily applied for
practical modeling of the extinction and scattering properties of
interstellar grains.
\end{abstract}


\section{Introduction}
Recently there has been considerable progress in the understanding of
interstellar dust. Perhaps the most pressing problem is the
uncertainty in the reference abundances (the total abundances of
elements in the gas {\em and} dust phases in the interstellar medium;
Savage and Sembach 1996; Mathis 1996). A reduction of the reference
abundance to about 70\% of solar now seems likely. These lower
abundances, coupled with the rather large amount of carbon in the gas
phase (Sofia, Fitzpatrick, \& Meyer 1998), make it very difficult to
understand all of the known properties of diffuse interstellar dust
while staying within the abundance limits. The abundance constraints
can be met better with grains that are composite (Mathis 1996),
containing various materials combined with small particles or voids
within porous grains, than by nonporous grains as envisioned by models
such as ``MRN'' (Mathis, Rumpl, \& Nordsieck 1977) or the very
important extension of MRN by Draine \& Lee (1984).

A major improvement in our understanding of dust has been brought
about by laboratory studies providing refractive indices for most
of the likely candidate materials (e.g., Schnaiter et al. 1998,
Mennella et al. 1998 for amorphous carbon). These studies must be
supplemented by a theory of determining the optical properties of the
composite grain, even if the optical constants of each constituent are
known. The problem is in determining the influence of the individual
components on each other at a microscopic level within the porous
grain.

Possible approaches include ``Effective Medium Theories," which try to
match the polarizability of the composite grain with a hypothetical
uniform medium (Bohren \& Huffman 1983, hereafter BH; Petrov 1986;
Ossenkopf 1991; Lakhtakia \& Thompson 1996; Ch\'ylek \& Videen 1998;
Videen \& Ch\'ylek 1998) with an index of refraction that is a
suitable superposition of the indices of the constituents. The problem
is how to average the indices, and in this paper we compare five
recipes with our solution. Other techniques for determining optical
properties are the ``Discrete Multipole Theories'', in which the
composite grain is replaced by an array of multipoles (often electric
dipoles only, for computational simplicity) in a specified
geometry. The computation determines how each multipole is affected by
the incident wave plus all of the other multipoles. This approach is
very general and can be used to determine how arbitrarily shaped
grains absorb and scatter light as well as such subtleties as the
effects of the sizes and shapes of voids within a fluffy grain (e.g.,
Wolff et al. 1994; Wolff, Clayton, \& Gibson 1998). However, it is not
practical to model large regions of parameter space (grain size,
wavelength, varying amounts of various likely grain constituents) with
present computational resources.

Our basic premise is that interstellar grains consist of either very
small homogeneous particles/molecules (sizes $\lesssim$ 10 nm) of a
particular composition, or else of larger grains with a
composite character. The composite grains are formed by
agglomeration of the very small grains, and smaller composites, within
dense regions. The composites produce most of the interstellar
extinction and polarization, and radiate their absorbed energy at mid-
and far-infrared wavelengths. They are chemically inhomogeneous, since
the individual particles they contain have differing chemical
compositions. Many modern grain theories make these assumptions (e.g.,
Mathis \& Whiffen 1989; D\'esert et al. 1990; Ossenkopf 1993;
Weidenschilling \& Ruzmaikina 1994, Mathis 1996).

This paper discusses determining {\em exact} cross sections for the
extinction and scattering of an approximate (but probably reasonable)
composite grain model. The method outlined here can easily be
generalized to include more grain candidates, including a composition
gradient within the grain. It provides the cross sections with far
less computation than discrete multipole theories.

The assumption is that a composite grain consisting of a random
collection of particles of various materials, plus voids, can be
replaced by a series of concentric spherical layers, each of which has
the optical properties of one material or vacuum. (Hereafter we will
consider vacuum to be a material.) Thus, the interactions of the
materials within the grain is taken into account.  Of course, we make
no claim that actual grains consist of concentric layers (including
voids!); the assumption simply mimics the interaction of small
particles after they agglomerate into a composite grain while in a
dense interstellar environment.

We do not restrict the individual layers of a particular material to
be ``thin'', in the sense that $2\pi\,l/\lambda<1$, where $l$ is the
thickness of the layer, because there is no theory at present to
predict the distribution of sizes of the individual particles that go
into each composite grain. This size distribution is set by the
shattering of grains produced in stellar atmospheres by collisions
within the ISM.

In \S2 we give a short description of the method of calculation, with
the formulae given in Appendix A. Results for various wavelengths,
including two types of amorphous carbons, are in \S3. There is
discussion and a summary in \S4.
\section{The Calculations}
\subsection{The General Solution}
The theory of interaction of a layered sphere is straightforward in
principle, since the eigenfunctions of the scalar wave equation are
very well known and the boundary conditions are rather simple (Kerker
1969; BH; Lopatin \& Sid'ko 1988; Wu \& Wang 1991). We are not aware
of the explicit display of the solutions in the astronomical
literature and have included an outline of the recursive algorithm we
used (Wu \& Wang 1991) in Appendix A. These solutions involve the same
Riccati-Bessel functions (with the same notation) that occur in the
two-layered sphere as given in the subroutine BHCOAT displayed in the
appendix of BH.

For illustrative purposes we will consider only composite grains with
three materials -- silicates, amorphous carbon (AMC), and vacuum. For
simplicity, we assumed that the volumes of each of the materials are
equal, as is approximately the case for silicates and AMC if the heavy
elements are in solar proportions. The 33\% vacuum is probably also
reasonable.

Each of the three materials is assumed to occur in consecutive
concentric spherical layers that form a ``shell". The whole grain
consists of a specified number, \nsh, of concentric shells, each with
the three layers of the materials. In order to determine the effects
of geometry, we permuted the order of the material layers within the
shells. We found that the order has an effect that converges rapidly
with increasing the number of shells (see \S3). We varied \nsh\ from 1
to 11, but \nsh\ = 3 gives an excellent match to the \nsh\ = 11 case.

Within each shell the radii of the layers of each material cannot be
equally spaced; since the volume fractions are specified, the
innermost layer must be relatively thicker than the outer. However,
the volumes of the shells can be chosen arbitrarily. If the shells are
given equal volumes, the radii of the various zones in the first shell
are relatively large: even with 11 shells (33 layers) the innermost
layer has 1/$33^{1/3}$ = 31\% of the grain radius, while the outermost
layer occupies only about 1\% of the radius.

With large values of the size parameter of the grain,
$x\:(=2\pi\,a/\lambda$, with $a$ the radius and $\lambda$ the
wavelength of the radiation outside of the grain), the code has some
numerical problems if the materials are very refractive. One set of
AMC constants (Schnaiter et al. 1998) produced numerical problems if
$kx>20$ ($k$ is the imaginary part of refractive index).  At this
large an $x$, we do not need to be concerned about these problems for
astrophysical applications because a substitute is
available (see \ref{emtvar}).

\subsection{The Rayleigh Solution and Associated Effective Medium
Theory} Farafonov (1999) has produced an expression for the Rayleigh
limit ($x\ll1$, $|m|x\ll1$, where $m$ is the refractive index) that is
useful in some cases. It has advantages over the general solution
because it can easily be generalized to any number of similar
ellipsoidal layers (i.e., with the same eccentricity). Unlike the
results for layered spheres, the polarization of aligned grains can be
computed.  Furthermore, there are a great number of astronomical
situations, especially at long wavelengths such as the mid-infrared
and beyond, where the Rayleigh approximation is an excellent one. Our
comparisons show good agreement with the general solution up to
$x\sim1$, or even somewhat larger.

The Farafonov formulae involve the number and materials of
the layers as well as the shape of the ellipsoid. We will quote the
expressions without derivation.

The interactions of a ``small" (Rayleigh) ellipsoid with
light can be described completely by the complex electric
polarizability, \alph.  The \alph\ along each principal axis,
$s\:(\equiv
x, y, z)$, of the ellipsoid surrounded by vacuum is given by (BH,
p. 145):
\be
\alpha_s=V\,(\varepsilon-1)[1+L_s(\varepsilon-1)]^{-1}\:,
\ee
\noi where $V$ is the volume and $L_s$ is the ``geometrical factor"
associated with the direction $s$. The values of these factors are
given in BH. For spheres, they are 1/3 for each axis.

The extinction cross section, $C_{\rm ext}$, is described by the
extinction ``efficiency factor", \qex, defined by $C_{\rm ext}=\pi
a^2\,Q_{\rm ext}$, and similarly for other cross sections (e.g.,
scattering). The absorption efficiency of the ellipsoid for radiation
with the electric field along the direction $s$ is then $Q_{{\rm
abs},s}=4x$\,Im\{\alph$_s/V$\} and the scattering by $Q_{{\rm
sca},s}=(8/3)\,x^4 |\alpha_s/V|^2$. In the Rayleigh limit, the
absorption
dominates the scattering because $x\ll1$, so the absorption is almost
equivalent to the extinction.

The $\alpha_s$ for $N$-layered ellipsoids is (Farafonov 1999)
\be
{\alpha_s} = V \frac{{\cal A}_2-{\cal A}_1}
                   {3 [({\cal A}_2-{\cal A}_1) L_s + {\cal A}_1]}\:;
                   \label{alph0}
\ee
$$
\left( \begin{array}{c}
{\cal A}_1\\ {\cal A}_2  \end{array} \right) =
\left( \begin{array}{cc} 1  &  L_s \\
     \tilde{\varepsilon}_N & \tilde{\varepsilon}_N (L_s-1)
                         \end{array} \right)
\prod_{j=2}^{N-1}
\left( \begin{array}{cc} (\tilde{\varepsilon}_j-1) L_s
+1  &
                         (\tilde{\varepsilon}_j-1) L_s
(L_s-1)/\tilde{V}_j \\
                         -(\tilde{\varepsilon}_j-1)\tilde{V}_j   &
-(\tilde{\varepsilon}_j-1) (L_s-1)+1
                         \end{array} \right)\times \nonumber
$$
\begin{equation}
\times \left( \begin{array}{c}
(\tilde{\varepsilon}_1-1) L_s+1
\\
-(\tilde{\varepsilon}_1-1) \tilde{V}_1
\end{array} \right)\:, \label{alph}
\end{equation}
\noi where $\tilde{\varepsilon}_i = \varepsilon_i/\varepsilon_{i+1}$ is
the relative dielectric constant ($\varepsilon=m^2$) of the layers
($j,\,j+1$) and $\tilde{V}_j=(a_j\,b_j\,c_j)/(abc)$, the ratio of
volume of \mbox{$j$-th} ellipsoid to the total volume of a particle.
The
$\tilde{\varepsilon}_N$ for the outermost layer, $j\:=\:N$, is the
dielectric constant $\varepsilon_N$.

An ``Effective Medium Theory" (EMT) is a very commonly used means of
estimating the optical properties of composite grains. The whole
purpose of an EMT is to predict the dielectric constant that produces
the same polarizability as the composite grain. EMTs are fast and
simple to use, since they provide the optical constants to be used in
an equivalent homogeneous grain. EMTs are, by construction , designed
to fit the absorption and scattering properties of the composite
grain, since they depend upon the electric polarizability. Other
optical effects (especially, the angular pattern of scattering) differ
from the extinction in their dependence on the amplitude and phases of
the radiation throughout the grain. The use of EMTs is especially
dangerous for these properties (see discussion in
Ch\'ylek \& Videen 1998; Videen \& Ch\'ylek 1998).

We can use Equation (\ref{alph0}) to define a ``Layered-sphere EMT,"
$\varepsilon_{\rm Lay}$. The relation between the dielectric constant
and
electric polarization, when $L_s = 1/3$ (i.e., spheres), is
\be
\frac{{\varepsilon}_{\rm Lay}-1}
     {{\varepsilon}_{\rm Lay}+2} =
     \frac{\alpha}{V}
               = \frac{{\cal A}_2 - {\cal A}_1}
                   {{\cal A}_2 + 2 {\cal A}_1};
\ee
\be
\varepsilon_{\rm Lay}=(1+2\alpha/V)/(1-\alpha/V)
               = {\cal A}_2/{\cal A}_1\:.
\ee

A program to calculate \alph\ is available at the sites that contain
the programs for the general layered sphere.
\section{Results}
We will display the results of various cross sections in a somewhat
nonstandard but perhaps useful way. Plots of \qex\ against $x$, where
$x\equiv2\,\pi a/\lambda$, may be less useful than another possibility
described below.

An important question is, ``What values of $x$ are most relevant at a
particular wavelength?" This paper does not try to produce a theory of
interstellar grains, but nearly all studies (e.g., Kim, Martin, \&
Hendry 1994; Kim \& Martin 1995) have suggested size distributions
something like that of MRN, with a grain size distribution
$n(a)\propto a^{-3.5}$. In this case, the extinction optical depth is
given by
\be
\tau_{\rm ext} = \int n(a)\,C_{\rm ext}\,{\rm d}a
\propto\int Q_{\rm ext}\, a^{-1.5} {\rm d}a\
=\int Q_{\rm ext}\, a^{-0.5} {\rm d}\,\ln a\:.
\ee
\noi For a given wavelength, $a$ is proportional to $x$. We have,
therefore, plotted \qa\ against $x$, using a logarithmic axis for
x. These plots give directly the integrand of the optical depth
integral (the product of the number of particles per unit size times
their cross section) if the distribution is like MRN.

The cross sections depend upon $x$ and the optical constants. To
illustrate the situations likely to occur in possible grain models, we
will present results using optical constants of actual grain
candidates: ``astronomical silicate" (Draine 1985) and either of two
types of AMCs: the ``Be\,1" tabulated in Rouleau \& Martin (1991), or
else the ``Jena AMC" (Schnaiter et al. 1996)\footnote{The indices are
available from the site ``http://www.astro.uni-jena.de/Group/Subgroups
/Labor/Labor/data/carbon/ar.lnk''. For a description of the database of
optical constants for astronomy see Henning et al. (1999).}. The
``Be\,1'' reflects the method of preparation of the AMC by Bussoletti
et al. (1987) in the laboratory. The two AMC's are very different,
reflecting the variations in the method of preparation. We considered
the range 1
\mic\ $\ge\lambda\ge$ 0.125 \mic, but the behavior of the cross
sections is well represented by \lam\ = 0.55 \mic\ (the $V$ band)
and 0.22 \mic, for which we present most of our plots and
discussion. (The results for the full range of wavelengths will be
considered in a future paper on much more detailed grain modeling.)
Let us write the index of refraction, $m$, as $m=n+i\,k$. For
(silicate, Be\,1 AMC, Jena AMC) at \lam\ = 0.55 \mic\ we used $n$ =
(1.72, 2.08, 2.04); $k$ = (0.030, 0.801, 2.23), respectively. At \lam\
= 0.22 \mic, $n$ = (1.87, 1.64, 0.875) and $k$ = (0.02, 0.52, 1.12),
respectively.  Note the relatively large values of $k$ for the Jena
AMC.
\subsection{Variations with geometry and number of shells}
\begin{figure}
\figurenum{1}
\plotone{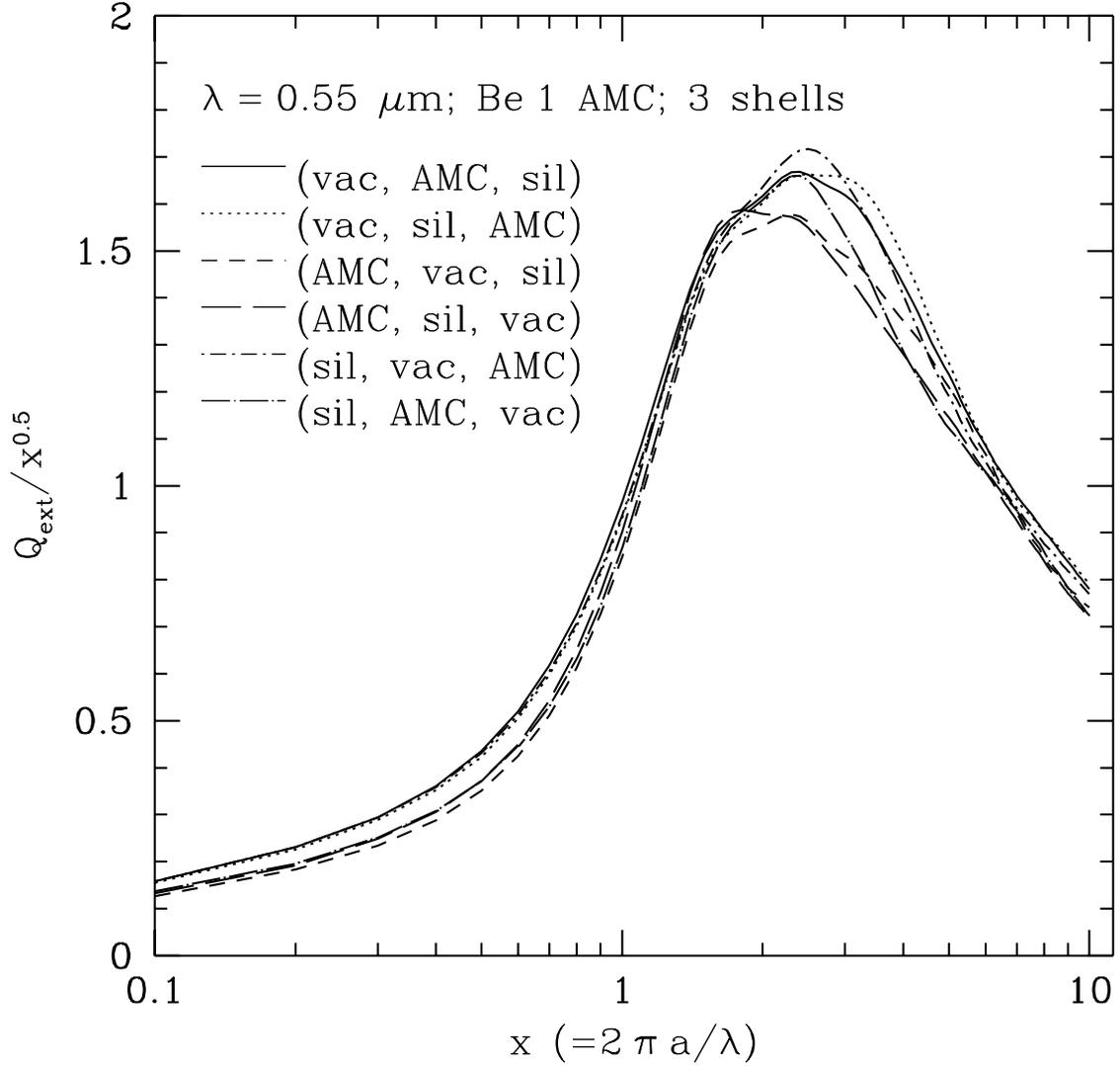}
\caption{\label{perms}
The quantity \qa, where $Q_{\rm ext}$ is the extinction efficiency of
a grain and $x=2\pi a/\lambda$, plotted against $x$ for the case of 3
shells, each containing equal amounts of vacuum, Be\,1 amorphous
carbon (Rouleau \& Martin 1991), and silicate. The constants are
appropriate for \lam\ = 0.55 \mic. The curves give the results for each
of the six permutations, as labeled. The quantity plotted is
proportional to the integrand of the extinction integral for a size
distribution proportional to $a^{-3.5}$, so the maximum of the curves
show the value of $x$ at which the contribution of grains is maximal.
}

\end{figure}

We first consider whether the order of materials within each shell
matters. Figure \ref{perms} shows \qa\ vs. $x$ for vacuum, silicate,
and ``Be\,1" AMC having equal volumes, \nsh\ = 3, \lam\ = 0.55 \mic,
and each shell having the same volume. The six curves represent the
permutations of the order of the materials within each shell.

We see that the \qa\ values vary by $\sim$ 25\% in the Rayleigh limit
and $\sim$ 15\% near the peak extinction. For almost all
values of $x$ the extinctions are largest when the vacuum is the
innermost material within a shell, since the absorbing materials are
then farthest from the center. Usually, the extinction is smallest
when the highly absorbing AMC is nearest the center, minimizing its
influence, and vacuum farthest, in which case the grain is actually
smaller. These trends show why fluffy grains may be more absorbing, per
gram, than nonporous ones: the farther the absorbing material
from the center, the better it can absorb, per gram.

A behavior similar to that in Figure \ref{perms} is shown when each
shell has an equal thickness rather than volume. In this case the
outer shells, which influence the absorption more than the inner ones,
are thicker that when the shell volumes are equal, and the cross
sections depend even more strongly upon the order of material than
shown in Figure \ref{perms}.

\begin{figure}

\newpage
\figurenum{2}
\plotone{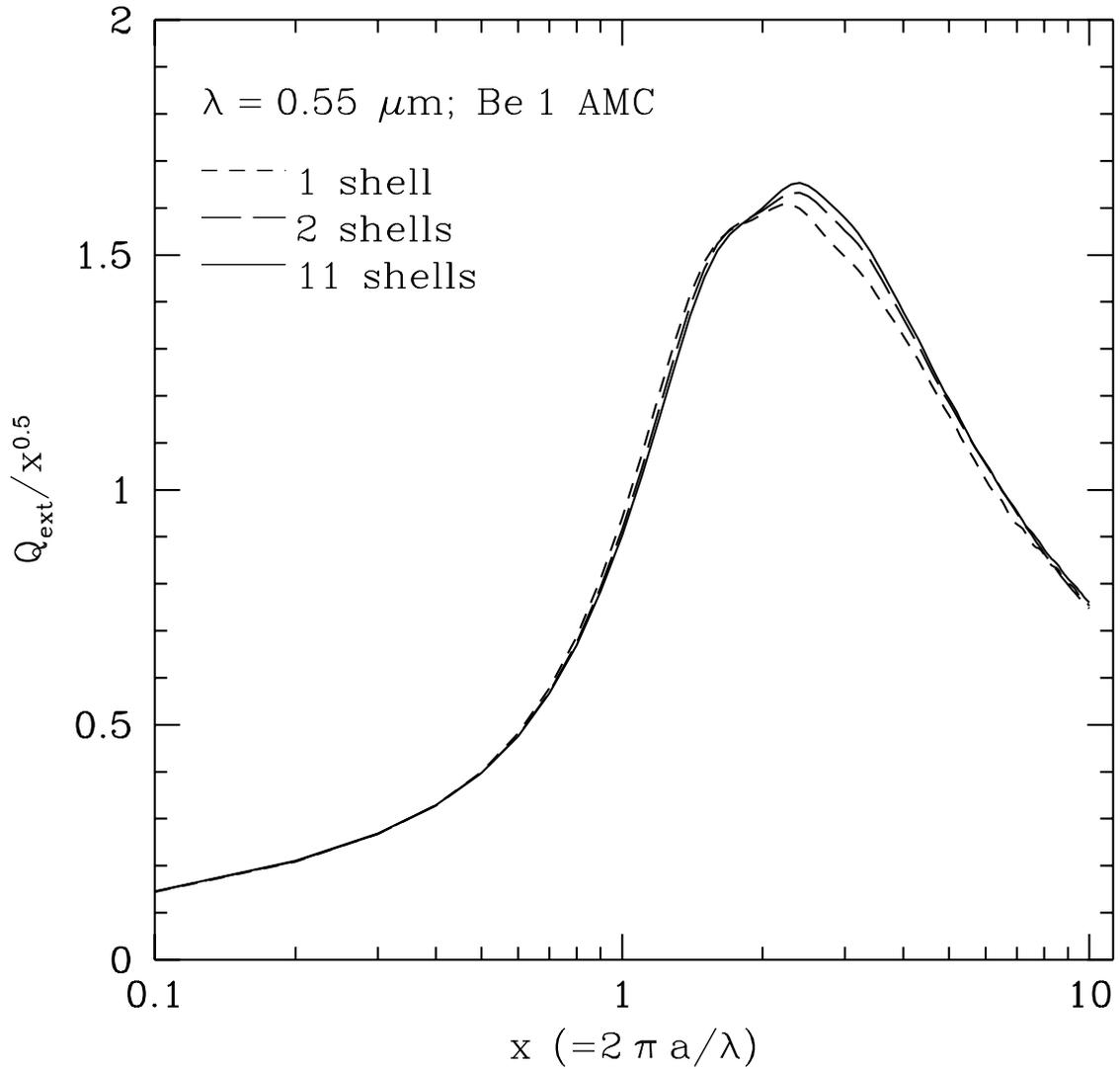}
\caption{
The quantity \qa plotted against $x$ with the cross sections averaged
over all six permutations of the order of materials within each shell
(see Figure \ref{perms}) for three values of \nsh. The optical
constants are the same as for Figure \ref{perms}. The dashed line,
dotted line, and solid line are for \nsh\ = 1, 2, and 11,
respectively. The lines for \nsh\ $\ge$ 5 are coincident with the
solid line.
\label{nshell}}  
\end{figure}

For larger values of \nsh\, permuting the order of materials becomes
unimportant. For \nsh\ = 11, the spread of the
six permutations (not shown) is about half that shown in Figure
\ref{perms} for three shells, with about the same averaged value
at each $x$. Figure \ref{nshell} shows \qa\ vs. $x$, averaged over the
6 permutations of material order, for \nsh\ = 1, 2, and 11. The
constants are the same as in Figure \ref{perms}. We see that there is
some deviation at \nsh\ = 1 from larger values of \nsh, but by \nsh\ =
2 the cross sections are very close to the values at larger
\nsh. The Rayleigh limit (small $x$) for the averaged \nsh\ = 1 is the
same as for larger values of \nsh.

For the remainder of this paper, we present results that are averaged
among the six permutations of the order of materials within each
shell, representing the physical situation of the materials and voids
being jumbled together within a composite grain.
\subsection{Comparison with Discrete Dipole Calculations}
The Discrete Dipole Approximation (DDA) has been used by a number of
authors (e.g., Draine \& Malhotra 1993; Henning \& Stognienko 1993;
Wolff, Clayton, \& Gibson 1998). These calculations determine the
response of an array of dipoles arranged in a cubic lattice with outer
boundaries that are as nearly spherical as possible. The strength of
each dipole is related to the index of optical constants of the
material it simulates.

The results from Wolff et al. (1998) are especially suitable for
comparison to multilayered spheres. They considered material similar
to silicate at optical wavelengths, with $m$ = 1.7 + 0.1\,$i$, along
with voids. They compared two cases: (a) individual dipoles were
removed randomly, so the sphere has fine-grained porosity; and (b)
each void had 20\% of the diameter of the original sphere, with voids
falling near the edge of the sphere producing cavities on the
surface. Two fractions of vacuum, 40 or 60\%, were considered. The
size parameters, $x$, were 1, 4, 7, and 10. In order to average over the
effects of the random removals, the calculations are repeated a number
of times.

In general, our multilayered spheres gave results intermediate between
the DDA with tiny voids (individual dipoles removed) and the large
voids (20\% of the original sphere's diameter), even in the limit of
very large \nsh.  An exception is for $x$ = 1, when the DDA results
were the same for large and small voids, and multilayered spheres gave
\qex\ 6\% larger for 60\% vacuum and 1\% smaller for 40\% vacuum. For
$x$ = 4, 40\% vacuum, the DDA found \qex\ = (3.20 $\pm$ 0.16) for both
large and small voids, where the ``error'' reflects the variations
found among various random locations of large voids. Our code gives 3.05
for \nsh\ = 50 and 3.15 for \nsh\ = 5. For 60\% vacuum, the DDA gave
\qex\ = 1.92 (small voids) and 2.15 (large voids); ours gives
2.08. Similar results hold for $x$ = 7. For $x$ = 10, the DDA for
small and large voids gave \qex\ = 2.85 and 2.5, respectively; ours
gives 2.59.

The physical reason why multilayered spheres do not agree with the DDA
with small voids is that our geometrical arrangement of the materials
is never really random and homogeneous. From the differences of the
DDA regarding large and small voids, it is clear that \qex\ depends
appreciably on the internal geometry of the grain. Our model contains
stacked layers throughout, and coherence effects will always be
present. On the other hand, there is no reason to believe that the
very fine voids modeled in some DDA calculations with very small voids
are physically correct. Almost surely real grains are rough
surfaced, irregular, and have their materials arranged in pieces that
may occupy an appreciable fraction of their size. Thus, there is
considerable uncertainty in any detailed model of grains.

Another comparison is with a four-material grain for which a DDA
calculation was kindly provided by M. Wolff (private communication,
1996). The materials are silicate, AMC, FeO, and vacuum, with
fractions of 0.421, 0.233, 0.046, and 0.3, respectively. The DDA was
provided for $x$ = 0.10, with the materials placed randomly in an
array of 113104 dipoles. For $x$ = 0.10, our code gave \qex\ = 0.083,
79\% of the DDA value.  The difference persists for large \nsh\ and is
not caused by the variations among the individual permutations of the
materials. Even in the Rayleigh limit, our grain model does not
represent arbitrarily fine inclusions. Probably real grains do not
either.

\subsection{EMTs and Variations with Wavelength\label{emtvar}}

We tested five EMTs: (a) the layered-sphere EMT (\S2.2); (b) the
``Bruggeman rule" (BH); (c) a modification of it, which averages over
grains of different shapes, perhaps simulating the complex structures
of layered spheres and real composite grains (Ossenkopf 1991;
Stognienko, Henning, \& Ossenkopf 1995)\footnote{With only spherical
grains, this Ossenkopf formulation provides the standard Bruggeman
rule.}; and (d) two versions of the Garnett Rule, often called the
``Maxwell Garnett Rule." This rule considers one material as a matrix
into which the others are embedded. We used either AMC or silicates as
the matrix.

Figure \ref{emt}{\em a} shows the accuracy of EMTs at \lam\ = 0.55
\mic. The AMC was Be\,1, and \nsh\ = 3. The solid line shows \qa\ for
layered
spheres, averaged over the permutations of the order of the materials.
The other lines are for the five EMTs as indicated in the figure. The
matrix for the Garnett EMT is noted.

Figure \ref{emt}{\em b} shows that EMTs are not always accurate in the
Rayleigh limit, as Figure \ref{emt}{\em a} might suggest. It shows the
results for a hypothetical material with ($n,\,k)$ = (1.4, 2.5) in
place of the AMC, along with ($n,\,k$) = (1.7, 0.03) for the silicate,
which are close to the real optical constants. The lines have the
same meaning as in Figure \ref{emt}{\em a}. By construction, the
Layered-sphere EMT is accurate in the Rayleigh limit, but the others are
not. The Bruggeman rule did rather well
($\sim$15\% too low). In general, the Ossenkopf rule does very well
for $x\lesssim$ 1 (9\% too low in this case) and is almost always
superior to the Bruggeman rule. Unfortunately, most of the extinction
integral occurs for $x>1$ for near-infrared and shorter wavelengths,
where the EMTs are not very accurate for the constants in Figure
\ref{emt}{\em a}.

Figure \ref{emt} shows that EMTs provide rather accurate extinction
cross sections at large $x$, where the layered-sphere code has some
numerical difficulties for very refractive materials. The EMTs use
homogeneous sphere calculations that are very stable and can be
extended to large values of $kx$, where our multi-layer code
encounters difficulties.

EMTs are often rather inaccurate near the maximum extinction. For some
sets of optical constants, one or more EMTs can be accurate near the
peak, but we find no simple relation that predicts which EMT is the
best, or how reliable it is.

Other wavelengths show results qualitatively similar to those in Figure
\ref{emt}, except that the peak shifts because of the changes in
optical constants. For \lam\ = 0.22 \mic, the peak occurs near $x$ =
2.8; for \lam\ = 0.125 \mic, it is at $x$ = 1.9. The EMTs predict the
peak position rather well but suggest too large an extinction near the
peak (see Figure \ref{emt}{\em a}), meaning that they give the
correct values of the real part of refractive index $n$ and
values of the imaginary part $k$ that are too small.

Our recommendation is to {\em use layered-sphere models for composite
grains, averaged over the permutations of the ordering of the
materials within each shell, with \nsh\ $\approx$ 3, in place of any
EMT.} If there is an assumed radial gradient of composition, a larger
\nsh\ is necessary, with the shells varying in composition, but there
is no problem in computing the cross sections in such a case.

The cross sections of the composite grains depend strongly upon the
optical constants used for the AMC. Figure \ref{jena} shows \qa\
vs. $x$ for both Jena and Be\,1 AMC constants, each with 33\%
silicates and vacuum, for 1.0 and 0.22 \mic. The curves are labeled in
the figure. We see that there is an appreciable difference in the
$\tau_{\rm ext}$ integral between Be\,1 and Jena AMC.
The Jena AMC curves stop at about $x=6$ because of
instabilities in the code at larger $x$. At 0.22 \mic\ the difference
of the materials is very large, and the Jena AMC will require
appreciably less carbon than Be\,1.

Figure \ref{jena} seems to show that $Q_{\em ext}$(1\,\mic) (solid
curve for Be\,1 AMC), integrated over the curves shown in the figure,
is larger than the integral of the dot-dashed curve (0.22 \mic). It is
well known that the opposite is true: the extinction at 0.22 \mic\ is
larger than at 1 \mic, even keeping in mind that we are not
considering the additional small particles that produce the 2175 \AA\
absorption feature. The paradox arises because the upper limits of the
integration are very different; they correspond to approximately the
same value of the grain size, $a$, and not $x$. The largest grain in
MRN is 0.25 \mic\ in radius, at which size the distribution is
abruptly truncated. If we assume, for illustration, that the volumes
of the largest composite grains are 70\% of the sum of the volumes of
the largest graphite and silicate grains in MRN, along with 33\%
vacuum, the radius of the largest composite grain is
$0.25\,[2(0.7)/(1-f_{\rm vac})]^{1/3}$ \mic, or 0.32
\mic. The values of $x$ for this size are 2.0 at 1 \mic\ and 9.25 at
0.22 \mic. Figure \ref{jena} shows that the upper limit of $x$ at 1
\mic\ will reduce the value of the integral of \qa\ over the sizes by
over a factor of two.  The inclusion of the gas-phase carbon, as well
as the carbon and silicon in the small grains or molecules producing
the red and infrared emission features, will reduce the upper bound on
$x$ from that estimated above, which will further reduce the 1
\mic\ extinction relative to the 0.22 \mic.

The actual size distribution in a model must be determined from a
careful consideration of all of the various demands for materials in
the light of all available observational constraints. For instance,
the integral over the composite grains does not include any of the
small carbon particles that provide additional absorption at 0.22
\mic, nor the carbonaceous molecules that produce the infrared emission
bands in the 3.28 -- 12 \mic\ region.

\subsection{Other Grain Properties}
Extinction is determined from observations far more accurately than
any other grain diagnostic property, but there are estimates (mainly
from reflection nebulae) of the albedo and of the phase parameter, $g$
$(\equiv\langle \cos (\Theta)\rangle$, where $\Theta$ is the angle of
scattering.) Gordon, Calzetti, \& Witt (1997) have given reasonable
estimates of these parameters as functions of wavelength. Any theory
using composite grains should make predictions for them as well as the
extinction per H atom. The same coefficients that are used in
calculating
the extinction cross sections also give the scattering cross section
and the phase parameter by standard formulae (e.g., BH). Here we
discuss the accuracy of the EMTs for providing the scattering cross
sections and phase parameters. We concentrate upon
\lam\ = 0.55 \mic\ because the optical constants there are very similar
to those throughout the optical region of the spectrum where most of
the observations exist. All of the results we present apply to this
region in general.

Figure \ref{qsca} shows $Q_{\rm sca}/x^{0.5}$ at 0.55 \mic\ plotted
against $x$. As for $Q_{\rm ext}$, this is the integrand of the
optical depth to scattering if the size distribution follows an
$a^{-3.5}$ distribution. The layered spheres consist of 33\% silicates,
vacuum, and each of the two AMCs. The solid line is the layered sphere
result (averaged, as usual, over all six permutations).

As expected, the scattering is very small until $x\gtrsim$ 1, even
with the size distribution assumed to be skewed to small grains
($\propto a^{-3.5}$). For the Be\,1 AMC, the EMTs overestimate the
scattering cross section because the interval of $x$ does not extend
much above 3 because of mass considerations; in this range the EMTs
are too large, and the Ossenkopf EMT is better than the Bruggeman. The
opposite is true for the Jena AMC; the cross sections are somewhat
lower than the layered sphere for $x>1$ and do not have the correct
limit at large $x$, where they are successful as regards
extinction. The Garnett rule with AMC as the matrix is the long-dashed
line that is far above the layered sphere near the peak scattering,
while the Garnett rule with silicate as the matrix is the closest to
the exact solution for Be\,1 AMC. Unfortunately, it was often the
worst in predicting the extinction (see Figure \ref{emt}). For Be\,1
AMC, the layered sphere lies below all EMTs for $x\le3$. From Figure
\ref{qsca} we conclude that EMTs are more unreliable for predicting
scattering than for extinction, and which EMT is suitable is not
clear.

The prediction of grain properties other than the absorption and
scattering cross sections (e.g., the angular distribution of the
scattering, or $g$) is especially dangerous because they depend upon
the details of the phases of the scattered wave within the grain, and
therefore on the details of the geometry. We will discuss $g$ while
keeping this danger in mind. Figure \ref{g} shows $g(x)$ for the same
parameters and notation as Figure \ref{qsca}, with Be\,1 AMC. For
$g(x)$ it is not appropriate to take the size distribution into
account because $g$ is not an averaged quantity, but the ratio of two
averages:
\be
g(x) = \frac{\int\,\cos(\Theta)\,Q_{\rm sca}\,{\rm d}\cos(\Theta)}
   {\int Q_{\rm sca}\,{\rm d}\cos(\Theta)}\:.
\ee

All EMTs predict almost the same $g(x)$, so only one line is plotted.
The envelope of deviations among the EMTs is typically $\sim$ 0.01, so
a superposition of the individual EMT results merely broadens the
dashed line slightly.

As expected, the figure shows that small grains scatter isotropically
($g\sim0$), and that large grains are forward-throwing ($g\ge 0.6$).
For the Be\,1 AMC, all of the EMTs do a poor job of predicting
$g$. Their prediction is somewhat better for Jena AMC, but the major
point is that in general EMTs are quite unreliable. However, the
assumption of spherical grains is also much more problematic for
predicting $g$ than for the extinction. For instance, spheres have a
strong back scattering from constructive interference of waves with
exactly equivalent paths at all azimuthal angles. This scattering is
not present in elongated or rough grains.
\section{Summary}
We have shown that layered spheres provide a simple way to calculate
the optical properties of spherical composite grains with an arbitrary
degree of vacuum and number of constituent materials, provided the
voids or individual constituent particles are small in comparison to
the composite grain (as is assumed in all EMTs as well). The variations
of the cross sections among the
permutations of materials for small numbers of layers (see Figure 1
for three shells, or nine layers) show that there are appreciable
effects of structure if the layers are thick in comparison to the size
of the grain.

\begin{figure}

\figurenum{3}
\plotone{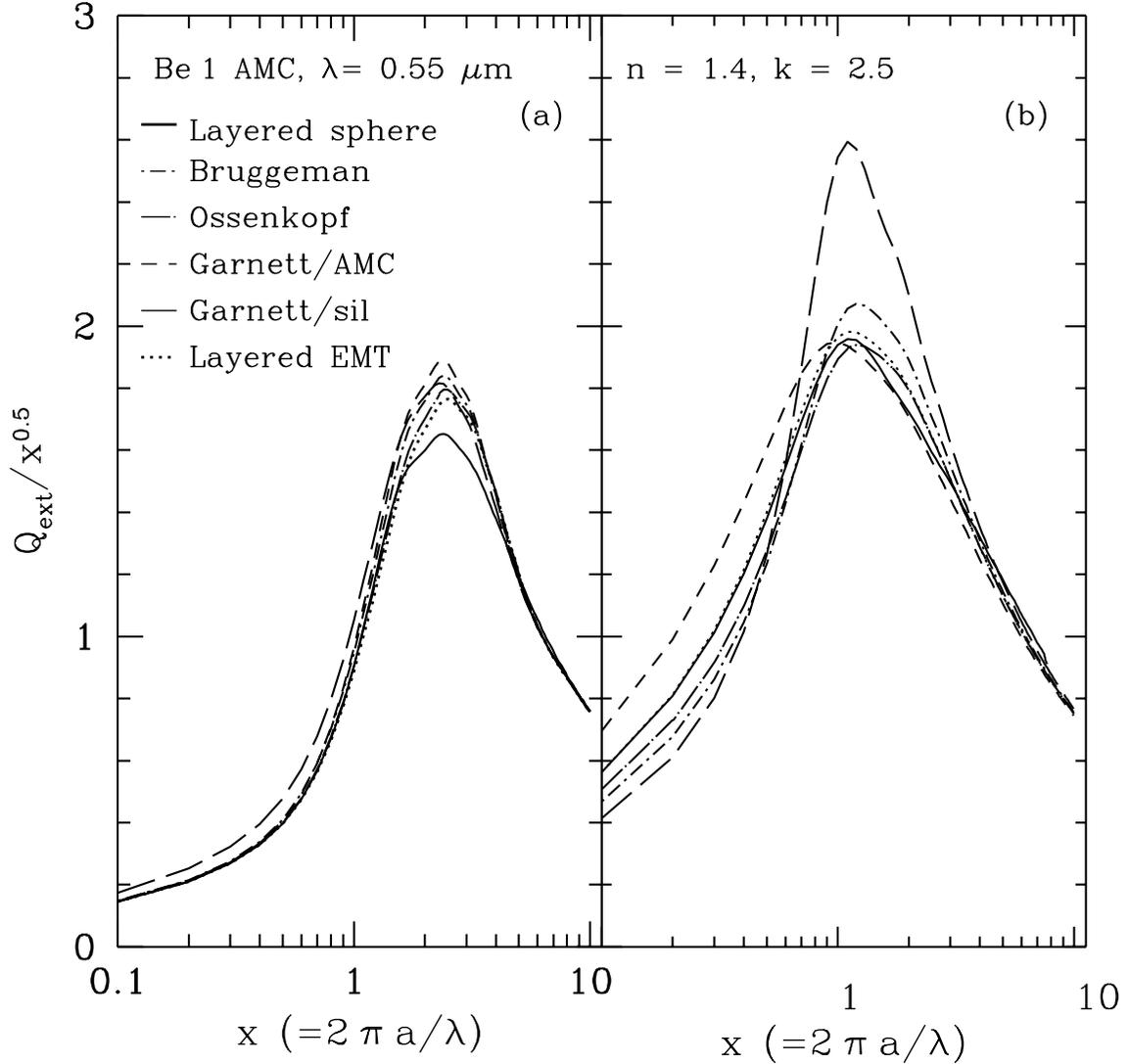}
\caption{
\qa\ plotted against $x$ with the cross sections from various
Effective Medium Theories (labeled) and also for the averaged over all
six permutations of the order of materials within each shell (see
Figure \ref{perms}). Vacuum, AMC, and silicate each comprise 33\% of
the volume.  The Garnett rule assumes one material, as distinguished
in the label, is the matrix in which the others are embedded. The
``Layered EMT'' is described in this paper.  The AMC constants are
({\em a}), Be\,1; ({\em b}) a hypothetical material with ($n,\,k$) =
(1.4, 2.5). We see that the EMTs uniformly overestimate the
cross section near the peak in ({\em a}) and are rather inaccurate at
small $x$ in ({\em b}).
\label{emt}}
\end{figure}

\begin{figure}

\figurenum{4}
\plotone{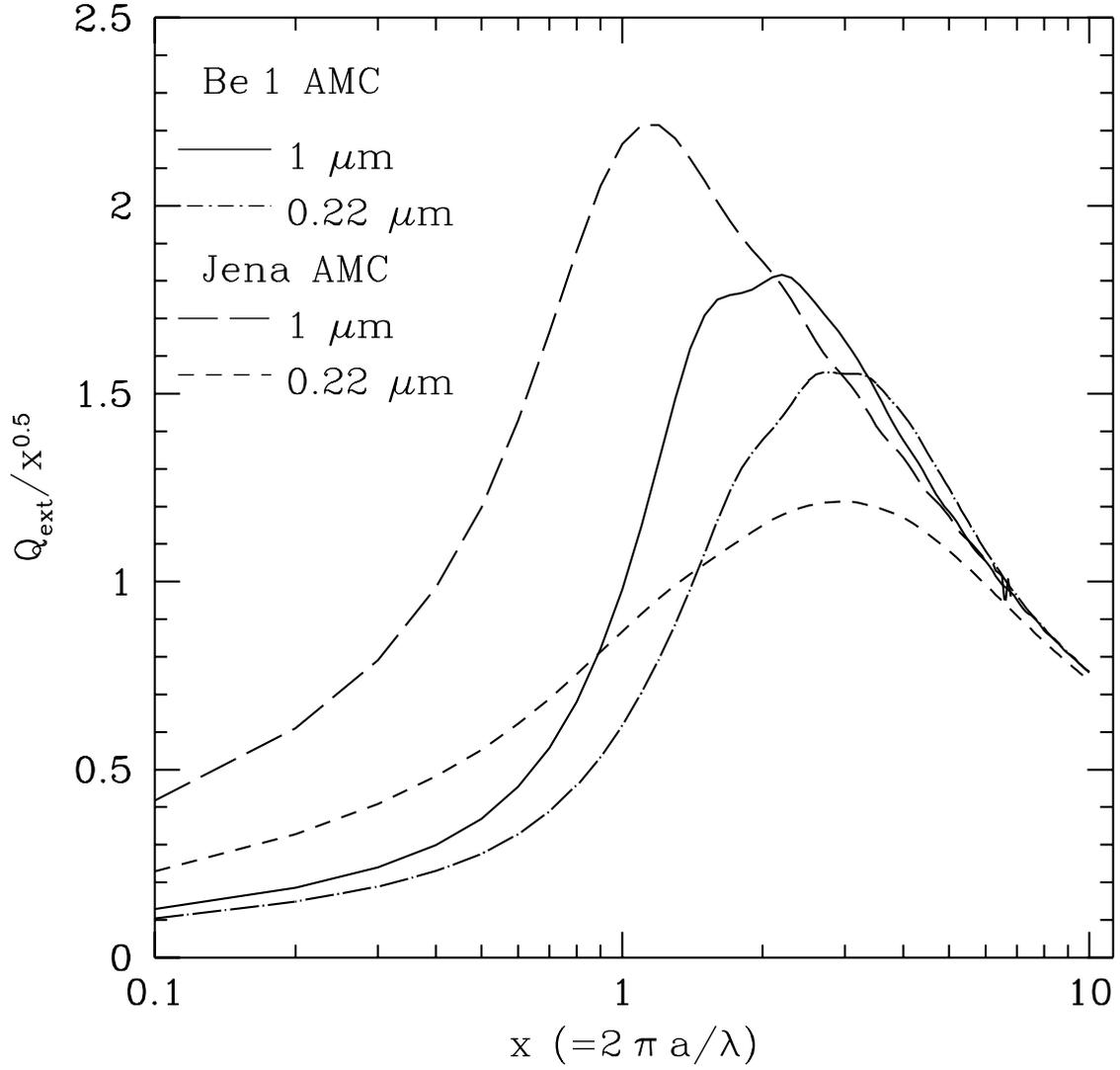}
\caption{
\qa\ vs. $x$ for two AMCs, two wavelengths, as labeled. Note the very
large difference in the cross sections for the two types of AMC at the
same wavelength.
\label{jena}}
\end{figure}

\begin{figure}

\figurenum{5}
\plotone{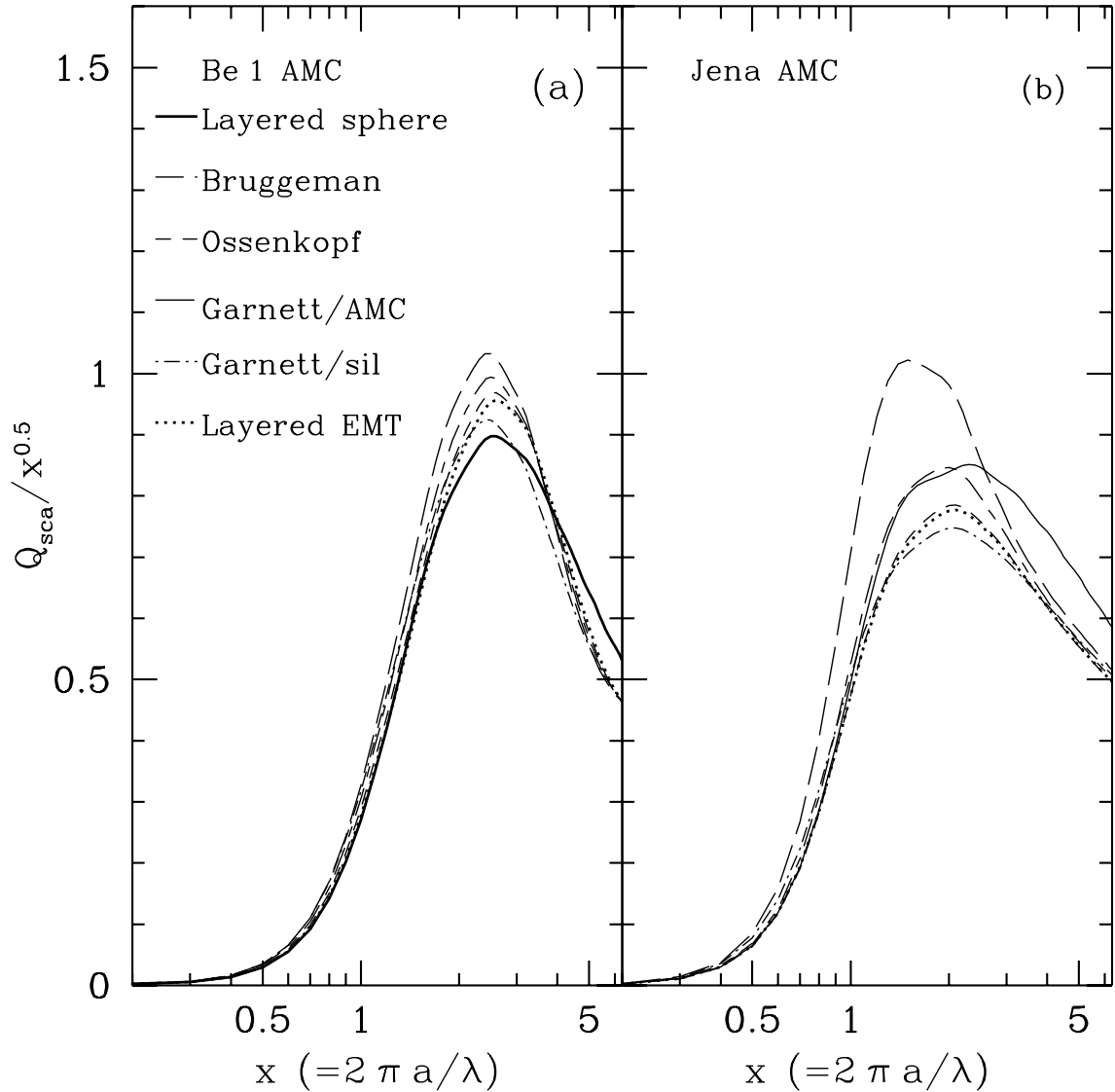}
\caption{
$Q_{\rm sca}/x^{0.5}$ vs. $x$, whose integral over $\log(x)$ gives the
optical depth for scattering, for layered spheres (33\% each of
vacuum, silicate, and amorphous carbon). The predictions of various
Effective Medium Theories, as indicated in the label, are also shown.
The wavelength is 0.55 \mic. In Figure ({\em a}) the AMC is Be\,1; in
({\em b}), Jena. The Garnett EMT with silicate as the matrix represents
the
scattering rather well, but it is worst for the extinction (see Figure
3).
\label{qsca}}
\end{figure}

\begin{figure}

\figurenum{6}
\plotone{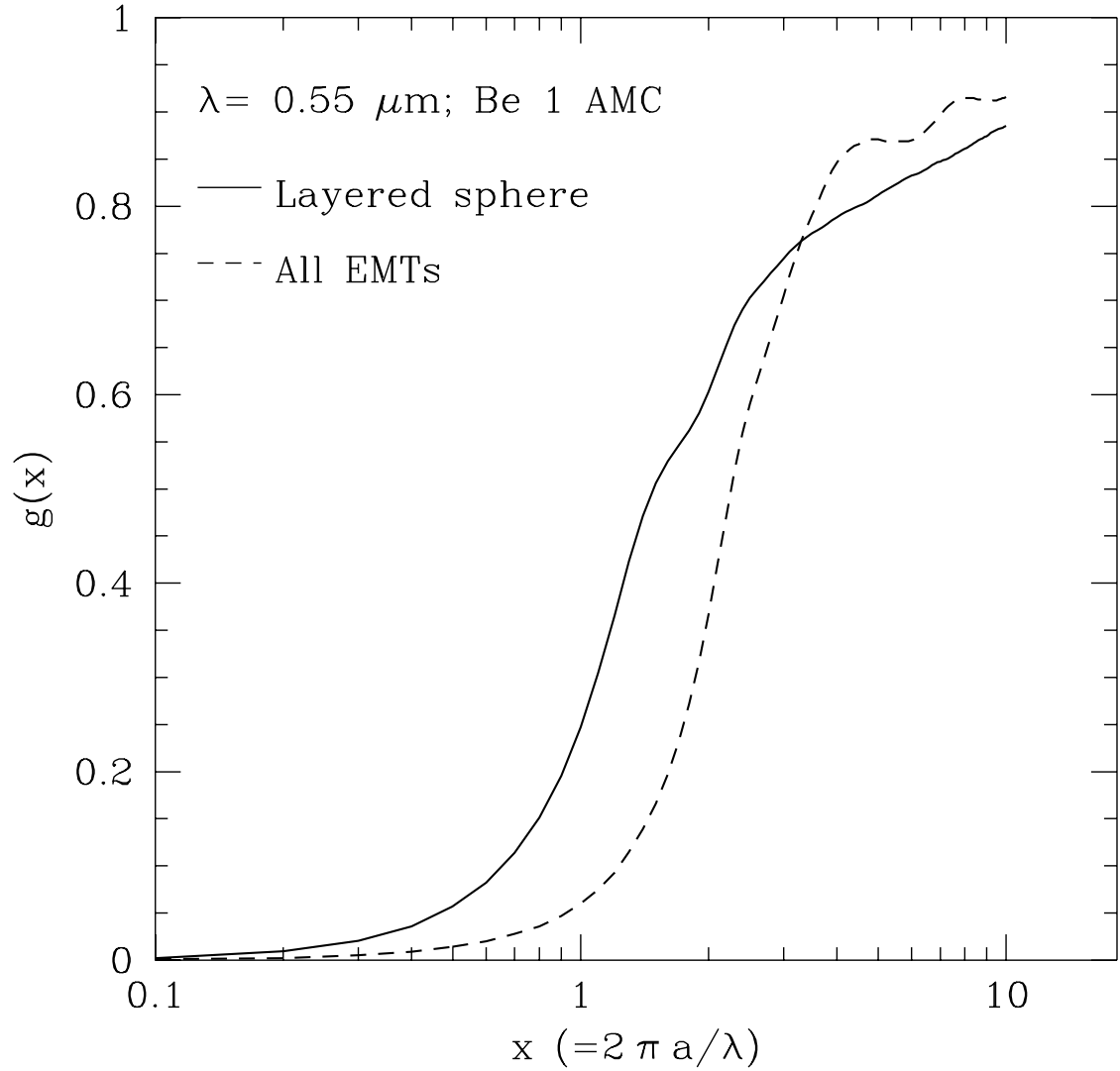}
\caption{
The phase parameter $g(x)\equiv\langle\cos(\Theta)\rangle$ for Be\,1
AMC, along with 33\% vacuum and 33\% silicates.  The solid line is the
layered sphere; various EMTs are identified in the label plotted
separately but are indistinguishable. We see that the prediction by
the EMTs is poor beyond $x\sim0.7$, especially for Be\,1 AMC, which is
less refractive than Jena AMC.
\label{g}}

\end{figure}

We have given an expression (Farafonov 1999) for the polarizability of
a layered sphere in the Rayleigh limit. This expression is quite
appropriate for wavelengths in the near infrared and longer. It can be
used to determine the optical properties of nonspherical grains,
including polarization introduced when they are aligned, because it
contains the shape dependence of the cross sections. We used it to
propose an ``Layered Sphere Effective Medium Theory'', in which an
average optical constant is taken to represent the ensemble of
materials. Our numerical results show that this EMT applies to
surprisingly large values of $x$ for predicting extinction and
scattering (Figures \ref{emt} and \ref{qsca}), but it fails, along
with all of the other EMTs, for predicting $g$ (Figure \ref{g}).

We have also shown that the various EMTs are quite approximate, with
the largest errors occurring for the scattering cross sections and,
especially, for the phase parameter, $g$. It is somewhat ominous that
the EMTs tend to overestimate the extinction cross section, since one
of the largest problems facing the understanding of interstellar
grains is meeting the cosmic abundance requirements, especially if the
reference abundance of the ISM is \mbox{sub-solar}.

The results of multilayered spheres do not represent composite grains
with extremely fine-grained constituent particles. There is always a
coherence in the assumed geometrical arrangement of the constituent
materials. Real cometary and interplanetary grains are also composed
of particles that can be appreciable in size as compared to the grain,
and probably interstellar grains as well.

The code producing the cross sections for layered spheres, in general
and also in the Rayleigh limit, is available from two sources: (a)
by anonymous ftp from ``ftp.astro.wisc.edu'' in the
directory ``outgoing/mathis''; (b)  via Internet:
``http: //www.astro.spbu.ru/JPDOC/nmie.html'' or
``http://www.astro.uni-jena.de/Users/database/nmie.html''.

\acknowledgments
We appreciate the suggestions made by the referee, Geoff Clayton.
N.V.V. was partly supported  by grants of the Volkswagen Foundation,
the program ``Astronomy'' of the government of the Russian Federation
and
the program ``Universities of Russia -- Fundamental Researches''
(grant N~2154). Both authors appreciate conversations with Prof. Th.
Henning.

\appendix
\begin{center}
{\bf Appendix A. Formulae for Multi-Layered Spheres}
\end{center}
We follow the formulation by Wu \& Wang (1991). The cross section
efficiencies (\qex, etc.) are given as series over an index $n$
involving two complex amplitude coefficients, $a_n$ and $b_n$ (see BH
or Kerker 1969 for explicit expressions.) We assume there are $N$
layers, with the \mbox{$j$-th} layer having a radius $a_j$, a size
parameter
$x_j=2\pi a_j/\lambda$, and a refractive index $m_j=n_j + k_j\,i$. The
outer radius is $a=a_N$ and size parameter $x=x_N$. The expressions
for coefficients $a_n$, $b_n$ can be written using the logarithmic
derivatives of Riccati-Bessel functions $\psi _{n}(z), \, \chi_{n}(z),
\,\zeta_{n}(z)$ and the ratios of one function to another, where $z$
is complex: $$ {\cal D}^{(1)}_n(z)\equiv
\frac{\psi_{n}'(z)}{\psi_{n}(z)} \:; \,\,\,\,\,\, {\cal
D}^{(2)}_n(z)\equiv \frac{\chi_{n}'(z)}{\chi_{n}(z)} \:; \,\,\,\,\,\,
{\cal D}^{(3)}_n(z)\equiv \frac{\zeta_{n}'(z)}{\zeta_{n}(z)} \:;  $$
$$ {\cal D}^{(l)}_n(z)= -\frac{n}{z}+\left[\frac{n}{z}-{\cal
D}^{(l)}_{n-1}(z)\right]^{-1}\:, \,\,\,\,\,\,l=1, 2, 3\:; $$ $$ {\cal
B}_n(z)\equiv \frac{\psi_{n}(z)}{\chi_{n}(z)} = {\cal B}_{n-1}(z) \
\frac{n/z+{\cal D}^{(2)}_{n}(z)} {n/z+{\cal D}^{(1)}_{n}(z)}\:; $$ $$
{\cal C}_n(z)\equiv \frac{\psi_{n}(z)}{\zeta_{n}(z)} = {\cal
C}_{n-1}(z) \ \frac{n/z+{\cal D}^{(3)}_{n}(z)} {n/z+{\cal
D}^{(1)}_{n}(z)}\:.  $$ The quantities ${\cal D}^{(1)}_n$ are
calculated by downward recursion in $n$, starting with ${\cal
D}^{(1)}_{N\!M\!X} = 1/z$ at the largest value of $n$, $N\!M\!X$. The
other functions above are calculated with forward recursion, starting
with $$ {\cal B}_0(z)= \psi_{0}(z)/\chi_{0}(z) = \sin z/\cos z \:; $$
$$ {\cal C}_0(z)= \psi_{0}(z)/\zeta_{0}(z) = {\sin z}({\sin z + i \cos
z})^{-1} \:; $$ $$ {\cal D}^{(3)}_0(z)= i \:.  $$ The ${\cal
D}^{(2)}_n(z)$ is best found from $$ {\cal D}^{(2)}_n(z)= \left[{\cal
C}_{n}(z){\cal D}^{(1)}_n(z)-{\cal D}^{(3)}_n(z)\right]\left[{\cal
C}_{n}(z)-1\right]^{-1}\:.  $$ Additional functions that vary with the
interface radii, $x_j$ ($j=2,...,N$), are defined in terms of the
above quantities: $$ {\cal H}^a_n(m_j x_j) =
\frac{{\cal B}_n(m_j x_j) {\cal D}^{(1)}_n(m_j x_j) - {A}^{(j)}_n
{\cal D}^{(2)}_n(m_j x_j)} {{\cal B}_n(m_j x_j) - {A}^{(j)}_n} \:; $$
$$ {A}^{(j)}_n = {\cal B}_n(m_j x_{j-1}) \ \frac{m_j {\cal
H}^a_n(m_{j-1} x_{j-1}) - m_{j-1}{\cal D}^{(1)}_n(m_j x_{j-1})} {m_j
{\cal H}^a_n(m_{j-1} x_{j-1}) - m_{j-1}{\cal D}^{(2)}_n(m_j
x_{j-1})}\:; $$ $$ {\cal H}^b_n(m_j x_j) = \frac{{\cal B}_n(m_j x_j)
{\cal D}^{(1)}_n(m_j x_j) - {B}^{(j)}_n {\cal D}^{(2)}_n(m_j x_j)}
{{\cal B}_n(m_j x_j) - {B}^{(j)}_n} \:; $$ $$ {B}^{(j)}_n = {\cal
B}_n(m_j x_{j-1}) \ \frac{m_{j-1} {\cal H}^b_n(m_{j-1} x_{j-1}) -
m_{j}{\cal D}^{(1)}_n(m_j x_{j-1})} {m_{j-1} {\cal H}^b_n(m_{j-1}
x_{j-1}) - m_{j}{\cal D}^{(2)}_n(m_j x_{j-1})} \:.  $$ The calculation
starts at the innermost interface, $j$ = 1, with $$ {A}^{(1)}_n =
{B}^{(1)}_n = 0 \:; \,\,\,\,\,\, {\cal H}^a_n(m_1 x_1) = {\cal
H}^b_n(m_1 x_1) = {\cal D}^{(1)}_n(m_1 x_1) \:.  $$ At the outer
boundary of the sphere the desired coefficients $a_n$ and $b_n$ are $$
a_{n} = {\cal C}_n(x) \ \frac{{\cal H}^a_n(m_N x) - m_N {\cal
D}^{(1)}_n(x)} {{\cal H}^a_n(m_N x) - m_N {\cal D}^{(3)}_n(x)} \:;
\,\,\,\,\,\, b_{n} = {\cal C}_n(x) \ \frac{m_N {\cal H}^b_n(m_N x) -
{\cal D}^{(1)}_n(x)} {m_N {\cal H}^b_n(m_N x) - {\cal D}^{(3)}_n(x)}
\:.  $$ The efficiency factors follow from these by a series $$ Q_{\rm
ext} = \frac{2}{x^2} \sum_{n=1}^{\infty} (2n+1) {\rm Re}(a_n + b_n)
\:;
\,\,\,\,\,\,
Q_{\rm sca} = \frac{2}{x^2}  \sum_{n=1}^{\infty} (2n+1)
(|a_n|^2 + |b_n|^2)  \:.
$$

\newpage

\end{document}